# Control of Intramolecular Electron Transfer in Perylene Dihydrazides and Perylene Diimides: A Comparative Study by Time-Resolved Spectroscopy

*Robin C. Döring[1], Eduard Baal[2], Malcolm A. Bartlett[2], Christian Prinzisky[2], Remco W. A. Havenith[3,4], Jörg Sundermeyer[2,\*], and Sangam Chatterjee[1,5,\*]*

[1]Faculty of Physics and Materials Sciences Center, Philipps-Universität Marburg, Renthof 5, D-35032 Marburg, Germany

[2]Faculty of Chemistry and Materials Sciences Center, Philipps-Universität Marburg, Hans-Meerwein-Straße, D-35032 Germany

[3]Zernike Institute for Advanced Materials, Stratingh Institute for Chemistry, University of Groningen, Nijenborgh 4, NL-9747 AG Groningen, The Netherlands

[4]Ghent Quantum Chemistry Group, Department of Inorganic and Physical Chemistry, Ghent University, Krijgslaan 281 (S3), B-9000 Gent, Belgium

[5]Institute of Experimental Physics I, Justus-Liebig-University Giessen, Heinrich-Buff-Ring 16, D-35392 Giessen, Germany

[\*]E-mail: sangam.chatterjee@physik.uni-giessen.de, jsu@chemie.uni-marburg.de

Electron transfer (ET) in molecular donor-acceptor dye systems is crucial for charge transport in organic semiconductors. Classically, ET rates should decrease with increasing donor-acceptor distance while the microscopic mechanism is more complex and shows intricate dependencies on the excitation conditions. In this paper, we introduce highly soluble *N,N'*-dialkyl perylene dihydrazides (PDH) – perylene dyes with a dialkylamino -$NR_2$ donor functionality directly bonded to both of their imide nitrogen atoms. We compare the PDH electron-transfer dynamics with a group of classical *N,N'*-bisalkylperylene diimides (PDI) equipped with a -$NR_2$ donor linked to the PDI acceptor core *via* a varying number of alkylene -$(CH_2)$- spacer groups, thus at distinctively different distance. Special physicochemical design features of our study objects include: i) amine moieties as donor group to minimize spin-orbit coupling; ii) substitution solely at both imide positions to avoid major impact on HOMO and LUMO levels and distortions of the PDI backbone; iii) control of donor-acceptor separation by non-conjugated alkylene groups to exclude any additional effects due to delocalized π electron systems. All materials show non-single-exponential photoluminescence decay dynamics. A rate equation analysis supported by electrochemical and absolute photoluminescence efficiency measurements yields evidence for efficient intersystem-crossing without heavy elements and reveals that the charge-transfer efficiency across the intramolecular interface strongly depends on the surplus excitation energy.

Perylene derivatives such as the perylene diimides (PDIs) where initially synthesized in the 1910's. From the 1950's on, they found a broad range of industrial dye applications, *e.g.* as high-grade pigments in industrial applications such as automotive finishes.[1] Since these early days, PDIs have developed into one of the most valuable material classes of organic electronics.[2-4] They have proven their value as versatile building blocks for functional optoelectronic supramolecular architectures.[5,6] Recently, advanced uses in contactless pH measurements, as metal cation sensors[7], and as active layers in organic photovoltaics (OPV) are proposed.[8]

Beyond their technological benefit, PDIs are furthermore model systems for charge and energy transport studies. Their high thermal and photochemical stability along with their preferable spectral range and electron acceptor properties favor such fundamental investigations.[6] For example, such systems provide insight into the time scales for charge generation by exciton dissociation and competing relaxation processes such as photoluminescence (PL), internal conversion (IC) or intersystem crossing (ISC). The intricate interplay between the different relaxation channels should be controllable by the chromophore type, its set of push (donor) and pull (acceptor) substituents and, in particular, by the supramolecular assembly of these organic semiconductor systems.

In general, perylene core substituents at either the bay- (1-, 6-, 7-, 12-) or the headland-/ortho- (2-, 5-, 8-, 11-) positions govern the triplet energy levels, the triplet quantum yields, and the potential to undergo carrier multiplication, *e.g.*, by singlet fission (SF). Recently, such multiple exciton generation is reported in slip-stacked thin-films of ortho-phenyl-substituted PDIs[9] and it is also expected to occur in rigidly linked dimers.[10] Rapid ISC with up to 53 % yield in heavy-atom-free, bay-substituted PDIs is observed in fluid solution and attributed to vibronically assisted spin-orbit coupling.[11] The intrinsic $S_1 \rightarrow T_1$ ISC efficiency of unsubstituted PDIs, however, is below 1 %.[3] Hence, its phosphorescence has only recently been reported in a glassy butyronitrile matrix containing methyl iodide at 77 K.[12] Consequently, triplet formation is not expected to play a major role in the excited-state dynamics of core-unsubstituted PDIs.

While a vast number of publications investigated core- and imide-*N*-substituted PDIs, only very few focus on perylene dihydrazides (PDHs). Early works use PDHs as fluorescence sensors for different organic targets.[13-15] More recent studies focus on nitrogen-nitrogen covalently linked multichromophore PDI systems. Utilizing a short fluorophore-fluorophore distance, nitrogen-nitrogen linked perylene and naphthalene imide dyads and triads are used as model systems for the investigation of (single-)molecular wires and intramolecular energy transfer.[16-19] Their use as dyes in p-type dye sensitized solar cells leads to a sharp increase of solar cell efficiency: the formation of dye-localized long-lived charge separated states enabled increased hole injection tendencies.[20-23] Unfortunately, planar PDIs are not yet able to compete with fullerenes as electron transporting materials despite their high electron mobility, thermal stability, and structural variety. This is mainly due to their strong tendency to form aggregates that reduces the overall charge separation and global transport.[24] Contrastingly, imide nitrogen-nitrogen linked PDHs result in perpendicular fluorophore orientation which leads to reduced molecular stacking and improved blend morphology in OPVs.[24] These beneficial properties spurr an increasing interest and make nitrogen-nitrogen linked PDIs promising alternatives to the fullerenes.[24-27]

In this paper, we present the hydrazinolysis of PTCDA with *N,N*-dialkylhydrazines leading to new and soluble PDHs. The focus is laid on the electron transfer across molecular-scale internal-interfaces between the amino moiety and the perylene core in PDIs and PDHs with direct nitrogen-nitrogen linkage. Three different routes to control the electron transfer rate are employed: i) variation of the orbital energies of the introduced donor moiety by means of protonation and methylation, ii) variation of the (electron) donor – acceptor distance, and iii) variation of the excitation energy used in the optical experiments. In all cases, we use time-resolved photoluminescence as a direct probe for the population decay of the lowest-energy bright transition $S_1 \rightarrow S_0$. A careful analysis of the transient luminescence dynamics reveals the presence of a shelving reservoir (SR) state interacting with the bright $S_1 \rightarrow S_0$ transition. This is manifested in intrinsic deviations from the commonly reported single-exponential photoluminescence decay.[28,29]

## Results and Discussion

The basis of these experiments are tailored samples which enable systematic investigations. We are able to obtain the smallest possible donor-acceptor distance at the imide position, *i.e.*, a direct nitrogen-nitrogen bond by using unsymmetrical *N,N*-dialkylhydrazines as donor-bearing moieties. Alkyl chains of increasing length control the spatial nitrogen-nitrogen separation and reveal the distance dependence of the photophysical properties. We chose electronically isolating and non-conjugationg bridge units instead of the commonly used phenyl groups. This eliminates the potential influence of their conjugated π-electron systems which, themselves, provide rich carrier dynamics. The imide position of the PDIs is selected for substitution due to the presence of a node in the MOs of the PDIs ground state and excited state.[5] This further reduces the orbital overlap and electronic interactions and, thereby, presumably also the rate of electron transfer.[5] Additionally, we are able to unravel the influence of processes competing with electron transfer such as IC or ISC.

Two series of samples are sysnthesized for this comprehensive study: PDHs with a nitrogen-nitrogen bond (no spacer) and dimethyl amino-alkylene-substituted PDIs with different alkylene spacer lengths, an ethylene ($C_2$), a propylene ($C_3$) and a hexylene ($C_6$) chain (Scheme 1). Sublimable **PDH-2** is obtained by the reaction of PTCDA with the commercially available *N,N*-dimethylhydrazine and the highly soluble **PDH-1** from *N,N*-dihexylhydrazine. The synthesis of unsymmetrical *N,N*-dialkylhydrazines is accomplished by a two-step reaction, first the nitrosylation of secondary amines, followed by lithium aluminium hydride reduction of the cancerogenic *N*-nitrosamines.[30] The synthetical procedure is described in the Methods Section and Supporting Information.

**Scheme 1.** Synthetic route and chemical structures of the studied **PDH-1** and **-2**, **PDI-1** to **-4**, the corresponding protonated **H₂PDH-1** and **-2**, **H₂PDI-2** to **-4** and methylated **Me₂PDI-2**[a].

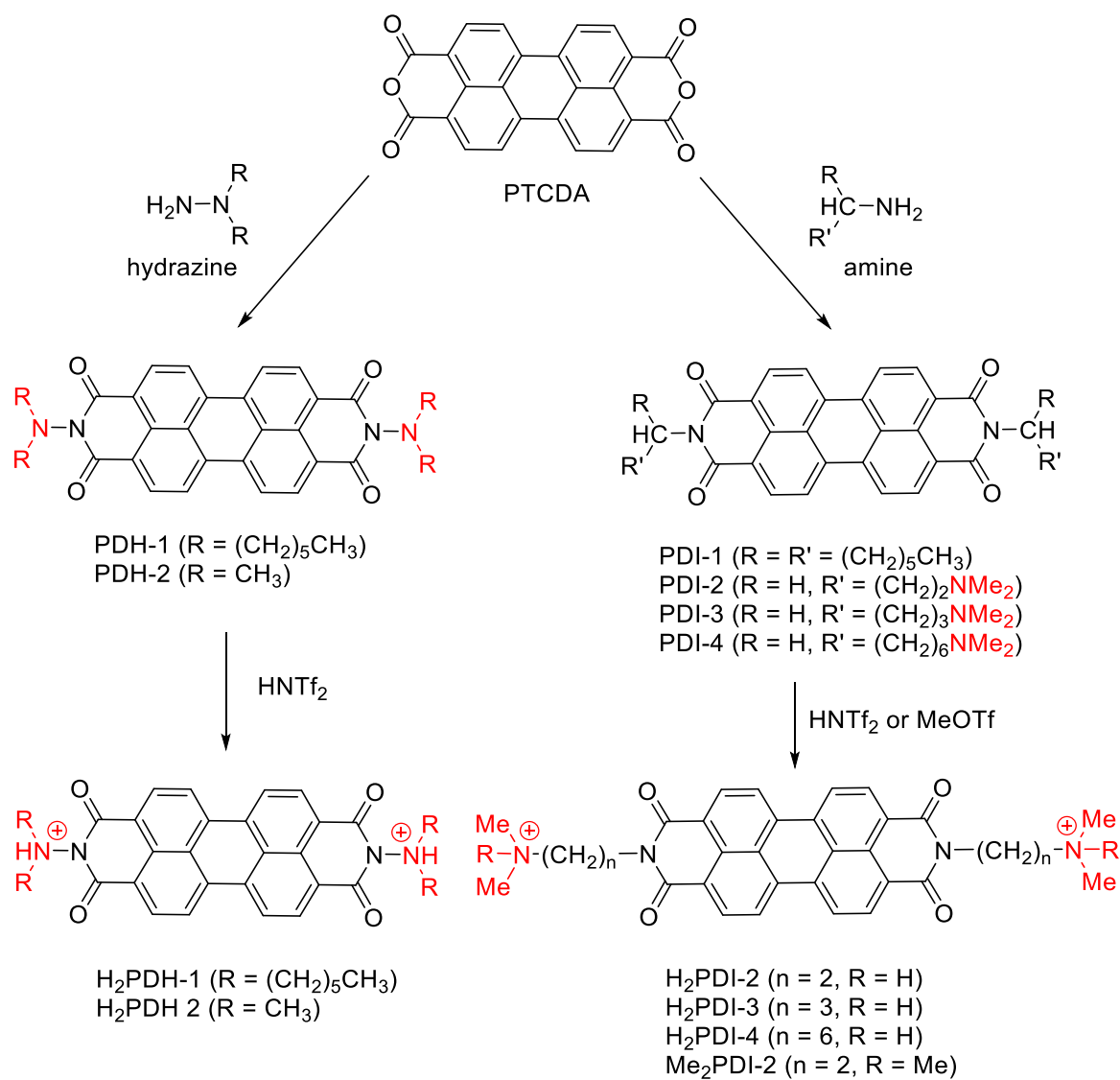

[a] Reagents and conditions: **PDH-1**, **PDH-2**, and **PDI-1 - PDI-4** are synthesized in molten imidazole at 130 °C and purified by column chromatography. Protonated **H₂PDH-1, H₂PDH-2,** and **H₂PDI-2** to **H₂PDI-4** are obtained from the corresponding PDH and PDI by reaction with bis(trifluoromethane)sulfonimide (HNTf₂) in toluene. Methylation of **PDI-2** in toluene yielded **Me₂PDI-2**. For detailed synthetical procedures see Supporting Information.

We present the crystal structure of **H₂PDH-2**, which consists of slip-stacked layers of the dicationic perylene units with an inter-layer distance of 11.53 Å within the unit cell. This relatively large distance guaranties only minimal fluorophore-fluorophore interactions even in a single crystalline material. The space between layers is filled with NTf$_2$ anions and solvents molecules (toluene). The shortest anion-cation distance is observed as a hydrogen bond between the oxygen of the NTf$_2$ anion and the acidic proton resulting in a three-dimensional network. Furthermore, an intramolecular hydrogen bond between the acidic proton and the carbonyl oxygen is found, which is confirmed by slightly longer carbonyl bonds for the participating carbonyl moieties (121.0(5) nm *vs.* 120.3(5) nm). This hydrogen bond locks the conformation and results in a perpendicular orientantion of the dimethylamino moiety to the perylene core and a minimalized steric repulsion. Such a hydrogen bond may, in general, cause a relativly low PL, since a carbonyl centered proton would lead to an unprotonated amine, which, in turn, enables electron transfer processes.

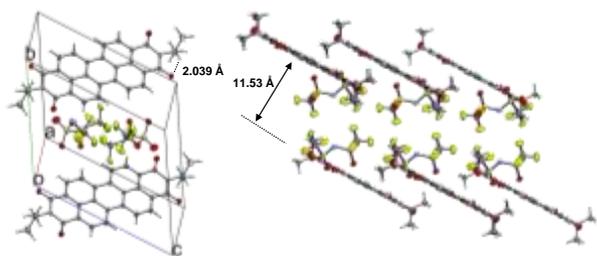

**Figure 1.** Crystal structure of **H₂PDH-2** showing the perylene cation and the NTf$_2$ anion. A hydrogen bond between the ammonium moiety and the anion is depicted. Solvent molecules are omitted for clarity. Crystals are obtained by slow evaporation of the solvent from a toluene-dichloromethane solution.

We follow this train of thought and investigate the influence of protonation on the PL of the individual molecules dissolved in organic solvents. Therefore, we turn to **PDH-1**. This new compound is related to the well-known reference compound, the "swallowtail" **PDI-1** of Langhals.[31] In **PDH-1**, a [=CH-] unit of PDI-1 is formally substituted by [=N-]. Both compounds show virtually identical absorption and emission spectra, *cf.* Figure. 2. However, the quantum yield, *i.e.*, the absolute external quantum efficiency of **PDH-1** is almost zero (<1 %), while **PDI-1** shows near-unity quantum yield. Upon protonation of the hydrazide –NR$_2$ group, the fluorescence shows slight indications of recovery. Doubly protonated dication **H$_2$PDH-1** is strictly isoelectronic to **PDI-1**. The protonation is expected to increase the oxidation potential of the donating amine, lowering the relative energy of the donor orbital, thus making electron transfer thermodynamically unfavorable. Indeed, the observerd minor increase of PL efficiency for **H$_2$PDH-1** strongly supports the previous assumption of the hydrogen bond leading to partial deactivation of the effect of protonation. Methylation, however, should enhance this effect as the bond between amine and methyl group is expected to be a lot more stable, virtually excluding any possibility of partial deprotonation in an equillibrium. Unfortunately, even the strongest methylation agent MeOTf did not yield *N*-methylated PDHs in refluxing 1,2-dichloroethane (12h, 80°C). Hence, neither **PDH-1** nor **PDH-2** are obtained as methylated species. (*cf.* Scheme 1). Instead we investigate a series of three related samples: the bare sample **PDI-2**, as well as readily protonated **H$_2$PDI-2** and methylated **Me$_2$PDI-2**.

We now turn the attention to the optical measurements. Generally, all samples under study have numerous spectroscopic properties in common: The spectral signatures, as obtained by absorption and steady-state fluorescence, are barely altered, *cf.* Figure 2a) and Figure 4a). In line with literature[32], the absorption maximum is located at 2.36 eV (526 nm) for all samples, when dissolved in chloroform (CHCl$_3$). The typical absorption-emission mirror symmetry is observed, owing to the Franck-Condon principle. Three vibronic progressions of the fundamental transition are observed, spaced equidistantly by ~170 meV. They are associated with a family of intramolecular C-C vibrations which couple to the optical transitions in these types of aromatic molecules. The relatively small Stokes shift of ~40 meV between the absorption and emission maxima indicates that the reorganization energy and changes in geometry are negligible for optical transitions. Hence, the perylene backbone governs the optical properties of the samples despite the chemical substitutions at the imide positions. This behavior is expected as the substituents are linked to the backbone in a non-conjugating manner. The influence of donor moieties linked by conjugated bridges have been extensively studied regarding their optical properties.[31-33]

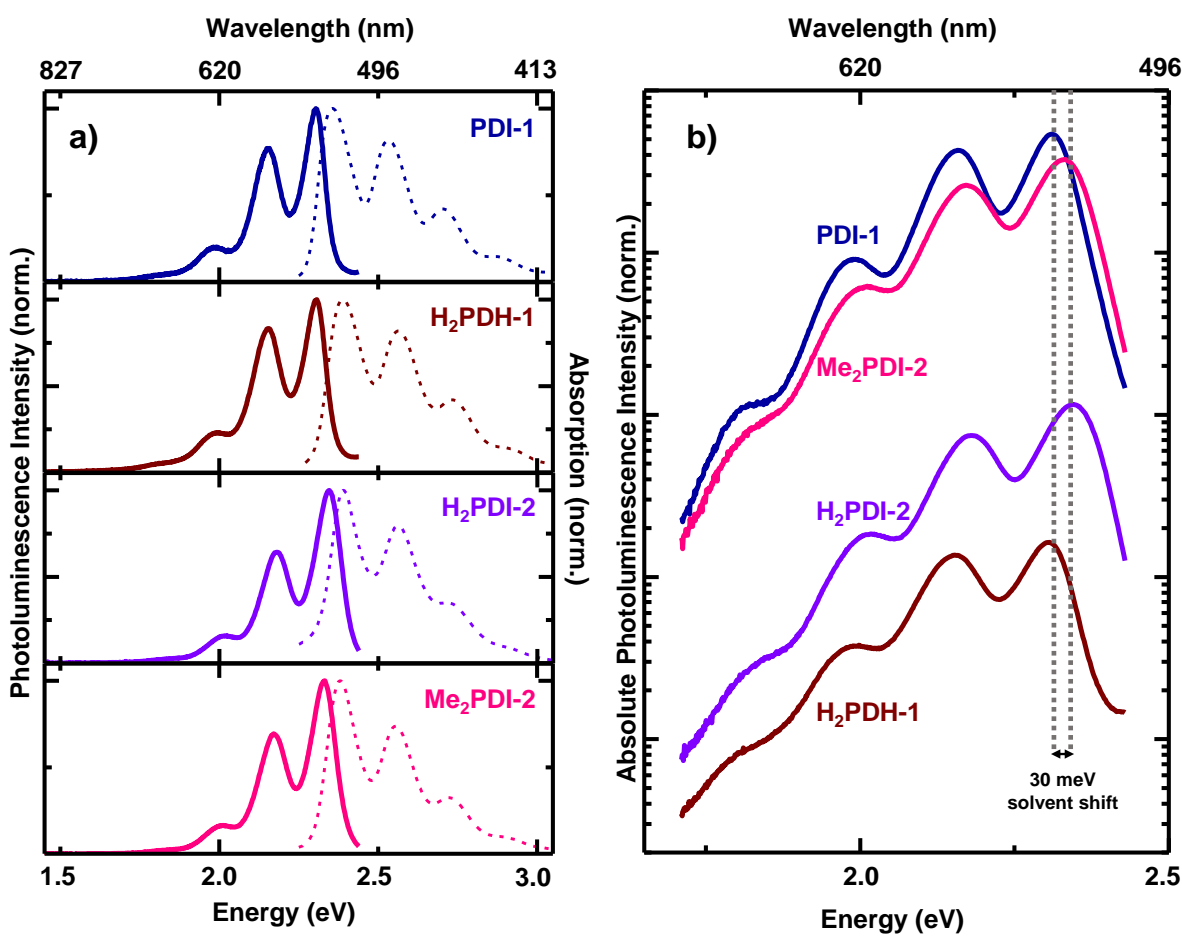

**Figure 2.** (a) Normalized absorption (dashed) and photoluminescence (solid) spectra of **PDI-1** (as reference), **H₂PDH-1, H₂PDI-2** and **Me₂PDI-2**. (b) Absolute photoluminescence of the same samples on a semi-logarithmic scale.

The minor spectral shift of 30 meV (*cf.* Figure 2b) for samples dissolved in acetonitrile (MeCN) is attributed to a change of the dielectric constant due to the higher polarity of MeCN, compared to CHCl$_3$. The quantum yields and lifetimes of **H$_2$PDI-2** only weakly deviate from **PDI-2**. This corroborates the assumption of partial depronotion which is also observed in **H$_2$PDH-1**. The quantum yield improves by a factor of two, accompanied by a lifetime increase of the fast component by a factor of roughly a (*cf.* Table 3). For the methylated compound, **Me$_2$PDI-2**, the situation is drastically changed. Its quantum efficiency almost matches that of reference **PDI-1.** The decay is single-exponential again and only about 10 % faster than that of **PDI-1**.

To shed further light on the observed deviation and identify the redox potentials we investigate the electrochemical properties by cyclic voltammetry (CV). The PDHs show two reversible reduction peaks similar to reference **PDI-1**. The first oxidazion of **PDH-1** is observed around 0.2 V lower than that of **PDI-1**. This is assigned to the oxidation of the amine moiety. Differential pulse voltammetry (DPV) reveals a second oxidation for **PDH-1** at 1.79 V, resulting in a $E_{Oxd,2} - E_{Red,1}$ of 2.46 V (see Figure S8), which is identical to the $E_{Oxd} - E_{Red,1}$ of **PDI-1**. Hence, we assign the second oxidation to the oxidation of the aromatic perylene core. This observation confirms the assumption that the GS of the amine is energetically lower than the HOMO of the aromatic core. Figure 3 shows the CV data of the reference compound **PDI-1** as well as those of **PDH-1** and **H$_2$PDH-1** in dichloromethane *vs.* SCE. **H$_2$PDH-1** and **H$_2$PDH-2** show only a single reduction peak each. A smaller $E_{Oxd} - E_{Red,1}$ (around 2.05 V) is found in acetonitrile due to a higher polarity of the solvent for both compounds. **H$_2$PDH-1**, which is also soluble in dichloromethane, shows a larger $E_{Oxd} - E_{Red,1}$ of 2.14 V in dichloromethane.

**Table 1**. Half-Wave redox potentials of PDHs and PDIs given in V *vs.* SCE.

| | $E_{Red,2}$ (V) | $E_{Red,1}$ (V) | $E_{Oxd,1}$ (V) | $E_{Oxd,2}$ (V) | $E_{Oxd,1-Red,1}$ (V) |
|---|---|---|---|---|---|
| **PDH-1** | -0.87 | -0.67 | 1.59 | 1.79[a] | 2.26 |
| **PDH-2** | -0.95 | -0.78 | 1.69 | - | 2.47 |
| **H$_2$PDH-1** | - | -0.12 | 1.92 | - | 2.04 |
| **H$_2$PDH-1**[c] | | | | | 2.14 |
| **H$_2$PDH-2** | | | | - | 2.07 |
| **PDI-1** | -0.84 | -0.64 | 1.82 | - | 2.46 |
| **PDI-1**[b] | -0.83 | -0.62 | 1.77 | | 2.39 |

Potentials are calculated with E(Fc/Fc+) = +0.46 V *vs.* SCE in dichloromethane and with E(Fc/Fc+) = +0.40 V *vs.* SCE in acetonitrile.[34] [a]Determined by DPV in dichloromethane. [b]Taken from literature.[35] [c]Values determined in dichlormethane.

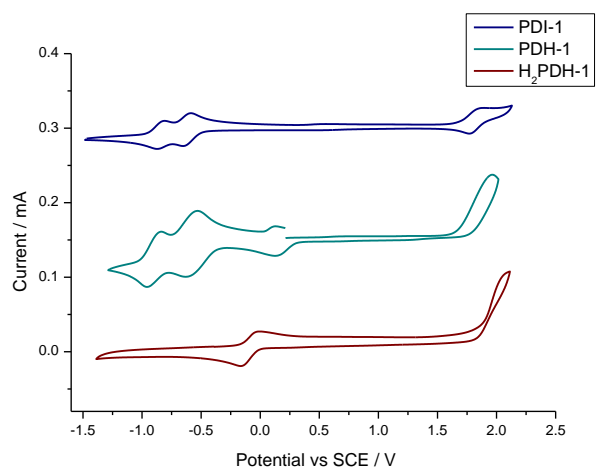

**Figure 3**. CV data of **PDI-1**, **PDH-1** and **H$_2$PDH-1** measured in dichloromethane versus SCE.

This assignment for the CV data is supported by DFT and TDDFT calculations for the neutral and doubly protonated compound **H$_2$PDH-2**. The results show that the MO associated with the terminal amine moiety is energetically intermediate to the aromatic π and π* orbtials of the perylene core. However, protonation stabilizes this MO to a level below that of the aromatic π orbital. These results also show how fluorescence quenching is energetically favored for the neutral compound but not for the protonated one. Figure 4 summarizes the results of these calculations.

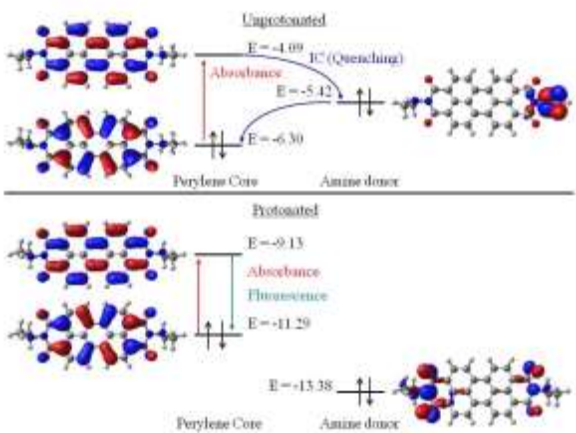

**Figure 4.** MO energy level diagram based on the S1 excited state B3LYP/6-31G* calculated for **PDH-2** and **H₂PDH-2** showing the effect of protonation upon the amine donor MO. Protonation stabilizes the amine MO and energetically favours fluorescence rather than internal conversion. Energies are given in eV.

Systematically varying the distance between the nitrogen and the amino donor moiety provides further insight into the electron transfer dynamics. Therefore, we use a series of samples with spacers introduced in between the nitrogen-nitrogen pairs of the reference compounds **PDH-1** and **PDH-2**. The spacers are alkyl chains of two, three and sixmethylene (CH$_2$) groups for **PDI-2**, **PDI-3**, and **PDI-4**, respectively.

The CV data allows us to extract the change in Gibbs free energy of the charge separation $\Delta G_{CS}$ using the redox potentials (complete data given in the Supporting Information) and the calculated donor-acceptor distances in the equation developed by Weller (eq 1).[36] The change in Gibbs free energy $\Delta G_{CS}$ is determined relative to the energy level of the PDI S$_1$ ($E_{0,0}$) which is obtained in the optical measurements.

$$\Delta G_{CS} = e\,(E_{ox} - E_{red}) - E_{0,0} - \frac{e^2}{4\pi\varepsilon_0\varepsilon_s R_{ee}} - \frac{e^2}{8\pi\varepsilon_0}\left(\frac{1}{r+} + \frac{1}{r-}\right)\left(\frac{1}{\varepsilon_{ref}} - \frac{1}{\varepsilon_s}\right). \qquad (1)$$

$E_{red}$ is attributed to the first reduction potential and $E_{ox}$ to the first oxidation potential. $E_{0,0}$ is the energy of the $S_1 \rightarrow S_n$ excited state, $R_{ee}$ is the edge to edge distance which corresponds to the nitrogen-nitrogen distance determined from DFT calculations. These values are only the upper limit of the donor-acceptor distance, since DFT calculations show stretched alkyl chains. $\varepsilon_{ref}$ is the dielectric constant of the solvent in the electrochemical measurements and is $\varepsilon_s$ the dielectric constant of the solvent used in the photophysical measurements (chloroform for neutral PDIs and acetonitrile for charged PDIs); $r^+$ and $r^-$ are the ionic radii of the amino cation and the PDI-anion (estimated 200 nm for the cation and 471 nm for the anion.[37]

**Table 2**. Estimates for the Gibbs free energy $\Delta G_{CS}$ for the formation of a charge separated state relative to $E_{0,0}$.

|  | $E_{0,0}$ (eV) | $R_{ee}$ (pm) | $-\Delta G_{CS}$ (eV) |
|---|---|---|---|
| **PDH-1** | 2.305 | 143 | 1.64 |
| **PDH-2** | 2.329 | 143 | 1.43 |
| **PDI-2** | 2.302 | 373 | 0.34 |
| **PDI-3** | 2.304 | 458 | 0.37 |
| **PDI-4** | 2.297 | 797 | -0.08 |

$\varepsilon$ (chloroform) = 4.81, $\varepsilon$ (dichloromethane) = 8.93, $\varepsilon$ (acetonitrile) = 37.5

The results of the calculation indicate that formation of CS becomes more favorable with decreasing donor-acceptor distances. While for **PDH-1** and **PDH-2** (no alkyl spacers) CS states are well below the energy of the excited PDHs. For $C_2$ and $C_3$ alkyl spacer bearing **PDI-2** and **PDI-3** the energy win is given but lower than for PDHs. In the case of **PDI-4** the CS formation even becomes energetically unfavorable. This formation of CS states suggests a PL quenching. Note that the Gibbs free energy for **PDI-4** is calculated assuming a stretched alkyl chain. Calculations using shorter donor-acceptor distances for **PDI-4** result in a more favourable CS with $\Delta G_{CS} = 0.14$ eV (for $R_{ee} = 500$ pm) and emphasize the importance of the donor-acceptor distance on the formation of the CS state.

These findings agree with the optical investigations. As shown in Figure 5, all samples show virtually identical absorption spectra, also similar to the first series shown in Figure 2. Spectral shapes and positions again show no effect on the appendage substituents. However, the quantum yields depend heavily an the distance between perylene backbone and donating amines illustrated in Figure 5b) which shows the absolute emission intensities of **PDH-1** and **PDI-1** – **PDI-4**. The addition of the amine groups and their electron lone electron pairs invoke a drop in PL intensity (and in quantum efficiency) of more than two orders of magnitude compared to **PDI-1**. The systematic introduction of alkyl chains as spacer units leads to a recovery of the quantum efficiencies. Eventually, **PDI-4** with a $C_6$ spacing unit reaches efficiencies of about 0.64, even in its unmasked neutral form.

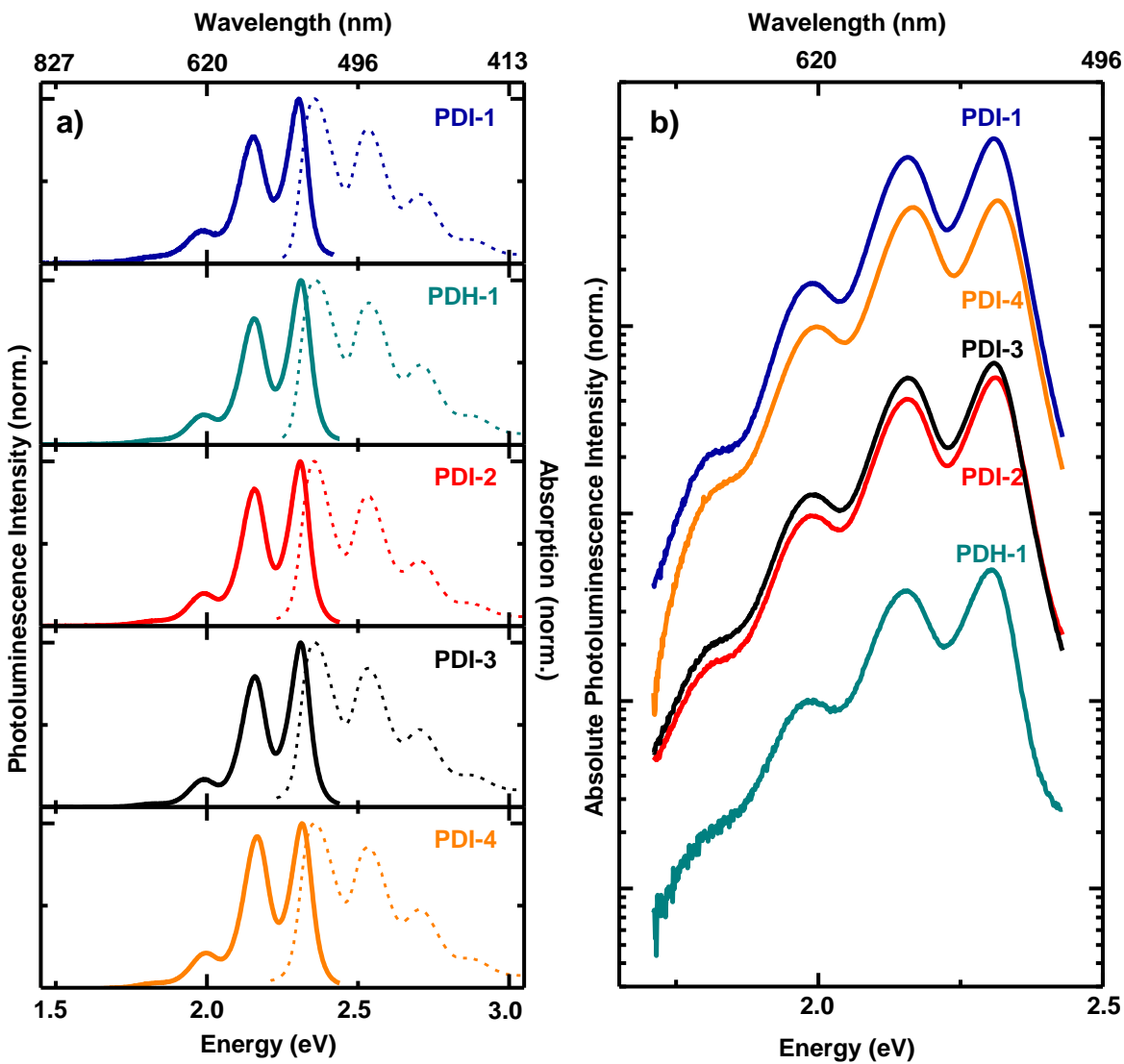

**Figure 5.** (a) Normalized absorption (dashed) and photoluminescence (solid) spectra of **PDI-1 – PDI-4** and **PDH-1**. (b) Absolute photoluminescence of the same samples on a logarithmic scale.

Time-resolved photoluminescence measurements provide further insight into the temporal dynamics of the population of the emissive excited singlet excited state (ESS). Characteristic decay profiles are shown in Figure S9. The reference compound **PDI-1** shows a single-exponential decay as expected due to its virtually exclusively radiative decay. Drastic deviations from this ideal behavior are found for the amino-substituted compounds. Samples **PDH-1**, **PDI-2** and **PDI-3** no longer show single-exponential behavior but rather more complex dynamics which are well-described bi-exponentially. While the initial decay ($t_{fast}$) is fastest for **PDH-1,** the lifetime recovers for larger distances and is almost single-exponential again for sample **PDI-4**, incorporating the longest ($C_6$) spacer.

The complex decay profiles of the substituted species include information on the electron transfer, which widely prohibits any radiative recombination from the $S_1 \rightarrow S_0$ transition. Decay profiles are fitted with an exponential (bi-exponential) function for the case of the unsubstituted (substituted) samples, respectively.

**Table 3**. Spectroscopic properties of the studied samples. Photoluminescence Quantum Efficiencies ($\Phi_{PL}$) and decay times measured by time-resolved photoluminescence.

|  | Solvent | $\Phi_{PL}$ ($E_{exc}$=2.8 eV) (%) | $t_{PL}$ ($E_{exc}$=2.8 eV) (ps) | $t_{PL}$ ($E_{exc}$=2.8 eV) (ps) |
|---|---|---|---|---|
| **PDI-1** | CHCl$_3$ | > 90.0 | 3907.9 ± 91.8 (= $t_{Ref}$) | |
| **PDH-1** | CHCl$_3$ | 0.5 | 17.6 ± 0.3 | 2735 ± 161 |
| **PDI-2** | CHCl$_3$ | 5.4 | 66.9 ± 0.4 | 3454.7 ± 129.9 |
| **PDI-3** | CHCl$_3$ | 6.1 | 150.4 ± 0.5 | 2710.8 ± 215.1 |
| **PDI-4** | CHCl$_3$ | 64.0 | 899.2 ± 11.7 | 3618.2 ± 158.4 |
| **H$_2$PDH-1** | CHCl$_3$ | 0.5 | 30 ± 1 | 2090 ± 528.7 |
| **H$_2$PDI-2** | MeCN | 10.3 | 254.4 ± 86.6 | 6368.4 ± 561.1 |
| **Me$_2$PDI-2** | MeCN | 89.0 | 3472.3 ± 84 | |

In accordance with previous reports, we assign the fast, initial quenching gives a figure for the rate of electron transfer: $k_{CS} = 1/t_{fast}$.[28,29,38] The decay profile for sample **Me$_2$PDI-2** could not be approximated by bi-exponential functions anymore. Here, the already low electron transfer rate is derived from the respective single-exponential lifetime ($t_{PL}$) and that of **PDI-1** ($t_{Ref}$)

$$k_{CS} = 1/t_{Pl} - 1/t_{Ref} \qquad (2)$$

Consequently, the quantum yields recovers along with the lifetimes, indicating that increased distance not only actually affects the electron transfer rate but also its efficiency.

The electron transfer rates are commonly described according to eq 3 within the framework of the Marcus picture which features the product of an electronuic term and a nuclear contribution:

$$k_{et} \propto \underbrace{|H_{DA}|^2}_{\text{electronic}} \cdot \underbrace{\exp\left\{-\frac{(\Delta G_{et}+\lambda)^2}{4\lambda k_B T}\right\}}_{\text{nuclear}} \qquad (3),$$

where $H_{DA}$ is the electronic coupling matrix element, and $\lambda$ the reorganization energy. Here, the electron transfer is solely governed by the electronic contribution $H_{DA}$ for $\Delta G + \lambda = 0$. To check this, we perform exemplary temperature-dependent measurements of **PDH-1** in a poly(methyl methacrylate) (PMMA) matrix (Figure S10). The spectral and temporal emission characteristics are temperature-independent in the range from 10-300 K, exactly as predicted by eq 3.

The plot of equilibrium edge-to-edge distance against $\log(k_{CS})$ (Figure S11) reveals a clear single-exponential dependence on the separation of donors and acceptors. The fit of the experimental data according to $k_{CS}(R_{ee}) = k_0 \cdot \exp(-\beta_{CS} R_{ee})$ infers that the involved mechanism is of the through-space type.[39] This appears plausible when considering the bridging units are alkylene chains with no delocalized π-electron systems. The electronic states of the spacer (bridge) lie too high in energy to be involved in the transfer process.

The observation of a non-single-exponential behavior for **PDH-1** and **PDI-3** – **PDI-4** can have various physical origins. The majority of which can be excluded due to careful measures taken during the experiments. Table 4 summarizes all possible non-intrinsic mechanisms, which lead to such an observed long-lived component. Dependencies on experimental conditions and observable effects respective mechanisms would show are also given. Additionally, the $^1$H-NMR spectra (Supporting Information Figure S1 to S4) rule out sample contaminations by another soluble PDI-specimen which then could provide the long-lived photoluminescence component. Aggregation as an explanation for a non-single-exponential behaviour seems unlikely for **PDH-1**, since a strong structural similarity to **PDI-1** is given.

Table 4. Non-intrinsic mechanisms leading to non-single-exponential decay dynamics.

| Mechanism | Dependency | | |
|---|---|---|---|
| | Sample Concentration | Excitation power | Spectral shift |
| **FRET** | ++ | o | + |
| **Two-Photon absorption** | o | ++ | o |
| **Multi-electron excitation (same or consecutive pulse)** | o | ++ | o |
| **Aggregation** | ++ | o | ++ |
| **Dimerization** | o | o | ++ |

"o": minor effect; "+": intermediate effect; "++": strong effect

The decay dynamics provides further insight into the charge-transfer dynamics of individual molecules as bimolecular or other aggregation effects and other extrinsic origins are excluded: varying the excitation fluence and concentrations in the sample solutions over several orders of magnitude provide identical results. Intriguingly, the PL decay is non-single-exponential for all samples. This infers the existence of an additional, reservoir, state in the molecule feeding the bright transition: the emission from an optical two-level system of localized states will always yield a single-exponential decay. Adding a second, possibly non-radiative decay channel again results in a single exponential decay. The combined decay rate in the law of decay is given by the sum of the two individual rates ($1/t_{combined} = 1/t_1 + 1/t_2$). Consequently, an additional reservoir needs to be involved to invoke a bi-exponential decay. A bi-exponential decay will result exclusively from the independent population of this reservoir, which is able to feed the originally considered emissive state. Taking into account this mathematical fact sheds new light on the obtained experimental results.

A rate equation model provides access to the effective interaction parameters with the reservoir state from the experimental data (Figure 6). We included four levels: the ground (GS) and excited states (ES) of the π electron system of the PDI core, the electron donating state (D) localized on the amine group – responsible for the luminescence quenching ($k_{CS}$) - and an electron reservoir state ($T_1$) interlinked with both, the excited ($k_s$ and $k_{ds}$) and the electron donating ($k_{dark}$) state, respectively. A state-filling term accounts for the Pauli-blocking of the molecular levels.

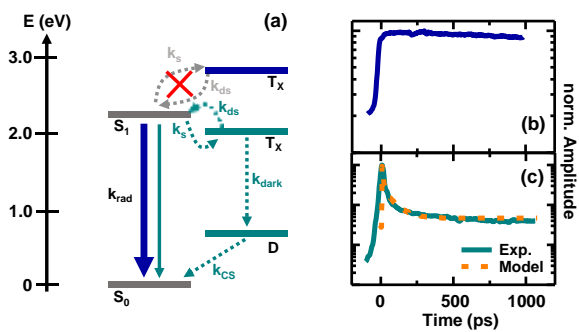

**Figure 6.** (a) Scheme of the energy levels and rate constants reference **PDI-1** (blue) and **PDH-1** (cyan). The weights of the arrows indicate the strength of the radiative transistion. The singlet and triplet geometries are also depicted. (b) Single-exponential decay observed in **PDI-1** for $S_1 < T_X$. (c) Non-single-exponential decay oberserved in **PDH-1** (solid line) and all other samples for which $S_1 < T_X$ is true. Also, a comparison to the model simulation (dots) is given, showing the good agreement between the two.

The model very well reproduces the experimental data and hence strongly suggests the existence of a reservoir state. However, its exact physical nature and the association of the transfer rates to physical processes are somewhat ambiguous. Regardless, the results clearly indicate that careful considerations are necessary whenever the photoluminescence decay of a molecular specimen deviates from the expected single-exponential behavior. For all samples showing bi-exponential decay, a shallow triplet state is found below the $S_1$ state; **PDI-1** forms the only exception, where the second triplet state is located at 2.67 eV (*cf.* Table S4). Hence, the shelving/deshelving process cannot occur and decay is single-exponential. We conclude the shelving/deshelving process actually constitutes a rapid forward and reverse ISC process.

Finally, we explore the dependence of the photoluminescence decay and, therefore, the electron transfer rate constants ($k_{CS}$) on the excitation photon energy. Four different excitation energies are used in time-resolved measurements, 2.5 eV, 2.8 eV, 3.3 eV, and 4.5 eV (*cf.* Figure 7a). Pumping at 2.5 eV and 2.8 eV excites the system into the $S_1$ electronic state, while the higher-lying singlet states, denoted $S_X$, are excited with excitation photon energies of 3.3 and 4.5 eV. The CS rates of all samples peak at 2.8 eV excitation energy. They subsequently decrease again for higher and lower excitation energies, respectively. **PDI-4** shows only a very weak dependence. This is probably owed to the overall low CS rate; the quantum yield already approaches that of **PDI-1**. The initial increase of $k_{CS}$ is most pronounced for **PDH-1**. For an excitation with 2.8 eV the system is transferred to a higher vibronic sublevel of the $S_1$ state (compared to $E_{exc} = 2.5$ eV). Thus, relaxation time from the excited vibronic sublevel to the emissive vibronic ground level is elongated. This leads to the observed increased $k_{CS}$ value, while at the same time PL intensity is reduced. Vibrational relaxation within the $S_1$ excited state of perylene in solution has been measured to occur on a timescale of 30 ps, very well on a timescale comparable to $1/k_{CS}$ of **PDH-1**.[40] For **PDI-2** – **PDI-4** $k_{CS}(E_{exc} = 2.5$ eV$)$ is lowered, compared to **PDH-1**. The dependence of $1/k_{CS}$ on the excited vibronic level within the electronic $S_1$ state is a lot less pronounced. To explain the successive decrease of $k_{CS}$ for $E_{exc} = 3.3$ eV and $E_{exc} = 4.5$ eV, we must pay attention to the behavior of donor-free **PDI-1**.

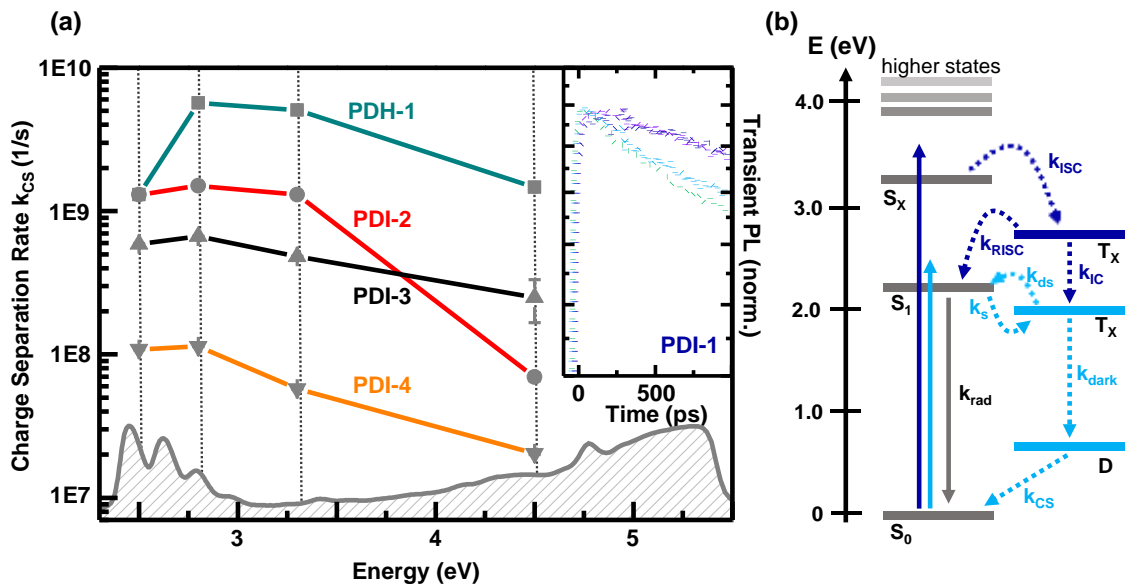

**Figure 7.** (a) Logarithmic plot of the Charge Separation Rate determined by TRPL measurements for samples **PDH-2** (squares), **PDI-2** (circles), **PDI-3** (up-pointing triangles) and **PDI-4** (down-pointing triangles). The dashed vertical lines indicate the respective excitation energies into the $S_1$ and higher energy $S_X$ absorption regions. *Inset*: Excitation energy dependence of the reference sample **PDI-1** plotted on a logarithmic scale. The color of the traces represent the color of the respective excitation wavelength. A clear increase of lifetime (fitting yields a 1.7-fold increase from 2.5 eV to 4.5 eV) with excitation energy is observed. (b) Extended energy level diagram showing the excitation (light and dark blue upward arrows) into and relaxation from the higher lying states ($S_X$, $T_X$).

No CS occurs for **PDI-1** (transient PL shown in the inset of Figure 7a). Nevertheless, the transients depend on the excitation energy, showing increased lifetimes for higher excitation energies. If a slow IC from $S_X \rightarrow S_1$ was the origin of the delayed fluorescence, the emission from $S_X \rightarrow S_0$ transition should be at least weakly observable. Such emission is demonstrated, *e.g.*, for matrix-isolated $SeO_2$ molecules.[41] Yet, we observe no higher energy emission for an excitation with 3.3 and 4.5 eV laser light. For pure perylene in liquid solution, the $S_3 \rightarrow S_1$ vibronic relaxation takes only as long as 60 ps.[33] Consequently, a slow IC alone is insufficient to explain a 1.7-fold lifetime-increase, assuming rates comparable to those observed in perylene. An ISC, however, is a possible explanation and is observed for similar systems, *i.e.*, in PDIs and NDIs lacking heavy atoms.[7,42] Significant ISC rates are observed especially for $S_x \rightarrow T_n (x > 1, n \geq 1)$ transitions.[7] Among others, a small singlet–triplet energy splitting is the key factor for efficient ISC.[7] According to DFT calculations, multiple triplet states exist around 3.2 eV for the studied samples (see Supporting Information Table S3). The photoluminescence excitation (PLE) spectrum deviates significantly from the absorption spectrum (Supporting Information Figure S12), indicating a partial breakdown of Kasha-Vavilov's rule. The deviations become particularly pronounced in the vicinity of the probed excitation energies. In the lower energy regions ranging from 2.2 - 3.0 eV, however, PLE and absorption data show no difference. This corroborates our assumption of an additional relaxation pathway, namely $S_x \rightarrow T_n (x > 1, n \geq 2)$ transitions. As the lifetime of the $S_1 \rightarrow S_0$ transition is increased for large $E_{Exc}$, the forward ISC must be accompanied by reverse ISC. Only the combination of both can lead to the observed 1.7-fold lifetime increase of the emissive state. The extended energy level diagram is shown in Figure 7b).

Strikingly, in the case of **PDI-2**, $k_{CS}$ at 4.5 eV excitation is lowered by more than one order of magnitude compared to the other excitation energies. An irreversible transition occurs under excitation with UV light accompanied by a drastic increase of fluorescence intensity. Mass spectroscopy revealed that the probable origin of this observation lies the cleavage of the bond connecting the PDI-core and the electron donating imide group. It remains unclear why a breach occurs exclusively for the sample bearing the $C_2$ spacer.

## Conclusions

In summary, we present the synthesis and characterization of two new PDHs, chromophores with a directly nitrogen bound donor –NR$_2$ attached to the acceptor perylene imide core. In particular, a well soluble **PDH-1 is** investigated, which is isoelectronically related to reference "swallowtail" **PDI-1**. These PDHs with very short donor-acceptor distances exhibit a very efficient fluorescence quenching. Upon double-protonation with bistriflylimide (HTFSI), the reported PDHs show partially recovered fluorescence. Such [H$_2$-PDH]$^{2+}$ salts reveal weak fluorophore-fluorophore interactions in solution and in the single-crystalline state as shown for bay-unsubstituted **[H$_2$PDH-2][TFSI]$_2$**.

The intramolecular electron transfer of these PDH is compared with a series of classical *N,N'*-bisalkylperylenediimides (PDI) equipped with an extra -NR$_2$ donor linked to the PDI acceptor core via a varying number of –(CH$_2$)- spacer groups, thus at distinctively different distances. Control of the transfer across the intralmolecular interfaces is achieved by three different methods. Donor methylation proved to be the most efficient, as indicated by quantum efficiency and time-resolved measurements. The donor-acceptor distances for the ground and excited state compounds are determined by DFT and TDDFT calculations, and used to calculate k$_{CS}$. The elongation of the spacer effectively prevents overlap of the donor and acceptor orbitals. The IC process responsible for fluorescence quenching in unprotonated compounds could be attributed to the MO associated with the terminal amine at an energy intermediary to the π and π* orbitals; protonation and/or methylation reduce the energy of this MO to below that of the π and π* orbitals, reducing the PL quenching.

The observed exponential distance dependence is in accordance with a through-space type mechanism. This experimental observation is unaffected by spurious effects in the spacer units from, *e.g.*, delocalized π-electron systems. The excitation energy dependence of k$_{CS}$ reveals the intricate interplay of charge-separation and IC processes. The pronounced non-single-exponential decay dynamics infer the presence of a dark shelving state within the system as extrinsic effects can be excluded. Furthernore, the excitation-energy dependent decay dynamics and bi-exponential nature of the transient PL strongly hint efficient ISC in these heavy-atom free, imide-substituted PDIs.

## Experimental Methods

**Synthesis.** Synthesis of **PDI-1** to **PDI-4** and **Me$_2$PDI-2** has been reported previously.[31, 43-45] Protonated PDHs and PDIs (H$_2$PDHs and H$_2$PDIs) are obtained analytically pure by protonation with bis(trifluoromethane)sulfonimide (HNTf$_2$) in toluene. The precipitated H$_2$PDH and H$_2$PDIsalts are characterized by NMR spectroscopy and elemental analysis. We chose the weakly coordinating TFSI anion to ensure a minimum of ionic interactions. Methylated **Me$_2$PDI-2** was obtained by reaction of **PDI-2** with methyl trifluoromethanesulfonate (MeOTf). Methylated PDHs are not accessible by reaction with methylation agents like methyl iodide, dimethyl sulfate or methyl triflate. Details of syntheses and characterisation are found in the Supporting Information.

**Electrochemistry.** Cyclic voltammetry (CV) and differential pulse voltammetry (DPV) of neutral compounds are performed in dichloromethane and of charged H$_2$PDHs and H$_2$PDIs in acetonitrile acetonitrile on a rhd instruments TSC 1600 closed electrochemical workstation (see supportion information). All samples are measured at a concentration of 5 mM or as saturated solutions for less soluble dyes in 100 mM tetrabutylammonium hexafluorophosphate solutions with a scan rate of 100 mVs$^{-1}$ with ferrocene as an internal standard. Half-wave redox potentials are summarized in Table 1 (a complete overview is given in the Supporting Information) and are given in V *vs.* SCE.

**Steady-state spectroscopy.** All steady-state absorption measurements are performed using a commercial double-beam spectrophotometer (Varian Cary 3) in the range from 190 – 900 nm. The steady-state emission spectra are obtained in a custom-built setup. A broad area diode laser operating at 2.8 eV (445 nm) is used for excitation and a compact spectrometer with 0.3 nm resolution (OceanOptics USB2000) for detection. The sample is mounted inside a 15 mm inner-diameter integrating sphere for the quantum-efficiency measurements. The detector is calibrated against a traceable tungsten-halogen white-light source; details are reported elsewhere.[46]

**Time-resolved spectroscopy.** A standard streak-camera setup is used for the time-resolved measurements. A pulsed titanium-sapphire laser (Spectra Physics Tsunami HP) acts as excitation source. Its sub-100-fs pulses are tunable between 690 and 1080 nm while operating at a fixed repetition rate of 78 MHz. The excitation wavelength range is further increased by including a frequency doubler and tripler unit. It provides the desired excitation energy of 2.8 eV (445 nm). An achromatic lens with f = 150 mm is used to focus the exciting laser beam on the sample. The emitted light is collected in backscattering geometry through the same lens and it is propagated towards a Czerny-Turner-type grating spectrometer yielding a spectral resolution of 2 nm. The streak camera equipped with a S20 photocathode that yields a time-resolution of 1.5 ps within the time window of 1.5 ns.

**Solution sample preparation.** The sample concentrations are kept well below the saturation limit for all samples and solvents (0.05 mM – 0.1 mM) to exclude aggregation effects, which would distort the optical spectra. The effect of oxygen induced fluorescence lifetime quenching is known to be negligible for **PDI-1** under air equilibration.[32] This assumption was verified by preliminary comparative measurements of $O_2$-free and air equilibrated samples. Hence, no further efforts are undertaken to avoid air equilibration. The samples are measured in standard cuvettes (Hellma Analytics) with an optical path length of 1 mm.

**Quantum Chemical Studies.** All calculations are performed using the Guassian09 software package.[47] The ground state (GS) structure for all compounds in both their protonated and unprotonated forms is optimized using DFT with the B3LYP functional[48–51] and the 6-31+G(d)[52,53] basis set. Frequency calculations are performed at the same level of theory as for the geometry optimizations; the absence of negative eigenvalues is considered as confirmation that the state obtained is a true global potential energy minimum and not a transition state. A full population analysis is then performed on the optimized GS structures using DFT with the B3LYP functional and cc-pVDZ[54] basis set. ESS structures are found by optimizing the GS structures using TDDFT with the B3LYP functional and 6-31+G(d) basis set. A full population analysis of these ESS structures was then performed at same level of theory as for their optimizations. Additional geometry optimisations (B3LYP/6-31G**) were performed using GAMESS-UK[55], and TDDFT (B3LYP/6-31G**) calculations for the first ten singlet and triplet excitation energies were performed using DALTON 2.0.[56,57]

# Author Information

**Corresponding Authors**
*sangam.chatterjee@physik.uni-giessen.de (optical spectroscopy, photophysics)

*jsu@chemie.uni-marburg.de (synthesis, structural, electrochemical analyses)

**Author Contributions.** JS initiated and created the idea to investigate isoelectronically related **PDH-1** and **PDI-1**. EB synthesized the samples, performed electrochemical measurements and analyzed the data; RCD and SC performed the optical experiments, analyzed the data using the rate equation model, and interpreted the results; CP provided the single-crystal structural data; MAB and RWH performed the DFT calculations. All authors contributed to writing of the manuscript. All authors have given approval to the final version of the manuscript.

## Acknowledgment

Funding by the Deutsche Forschungsgemeinschaft through SFB 1083 is gratefully acknowledged. C.P acknowledges funding by DFG through GRK 1782. M.A.B. acknowledges funding by a DAAD doctoral fellowship.

## Supporting Information Available

Details on the synthetic procedures, NMR spectra, electrochemisty TRPL, temperature- and distance-dependence of the electron transfer, details on the rate equation model, PLE data and DFT calculated singlet and triplett energies.This material is available free of charge *via* the Internet at http://pubs.acs.org.